\documentclass[review]{elsarticle}

\usepackage{amsmath,mathtools}
\usepackage{graphicx}
\usepackage{color}
\usepackage{txfonts}
\usepackage{units}

\journal{Physica A}

\newcommand{\ck}{c^{\vphantom{\dagger}}_{\! k}}
\newcommand{\ckd}{c^{\dagger}_{\! k}}
\newcommand{\ckp}{c^{\vphantom{\dagger}}_{k^{\prime}}}
\newcommand{\ckpd}{c^{\dagger}_{k^{\prime}}}
\newcommand{\cqd}{c^{\dagger}_q}
\newcommand{\cqpd}{c^{\dagger}_{q^{\prime}}}
\newcommand{\e}{\mathrm e}
\newcommand{\EjN}{E_{\! j, N}}

\newcommand{\epsk}{\epsilon^{\vphantom{\dagger}}_k}
\newcommand{\epskp}{\epsilon^{\vphantom{\dagger}}_{k^{\prime}}}

\newcommand{\fbraket}[1]{\mathinner{\langle{#1}\rangle}}
\newcommand{\fket}[1]{| #1 \rangle}
\newcommand{\FN}{F_{\! N}}
\newcommand{\Gt}{\tilde{G}}
\newcommand{\HNpart}{\hat{\mathsf H}_{N}}
\newcommand{\Hop}{\hat{\mathsf H}}
\newcommand{\im}{\mathrm i}
\newcommand{\kB}{k_{\mathrm B}}
\newcommand{\ket}[1]{\mathinner{\left| #1 \right \rangle}}

\newcommand{\fketbra}[2]{\mathinner{| #1 \rangle \langle #2 |}}
\newcommand{\muGCE}{\muup_{\mathrm{GCE}}}
\newcommand{\muN}{\muup_N}

\newcommand{\nkop}{\hat{n}^{\vphantom{\dagger}}_k}
\newcommand{\nk}{n^{\vphantom{\dagger}}_k}
\newcommand{\Nop}{\hat{\mathsf N}}
\newcommand{\ns}{n_{\mathrm S}}
\newcommand{\phik}{\phi_k}
\newcommand{\phikc}{\phi^*_k}
\newcommand{\phikp}{\phi_{k'}}
\newcommand{\psid}{\psi^{\dagger}}
\newcommand{\PN}{\hat{\mathsf P}_{\! N}}

\newcommand{\SN}{S_{\! N}}
\newcommand{\Sp}{\mathsf{Tr}}
\newcommand{\Stwo}{S_{\! 2}}
\newcommand{\Sfour}{S_{\! 4}}
\newcommand{\UN}{U_{\! N}}

\newcommand{\rhoN}{\hat{\rho}_N}
\newcommand{\FH}[1]{F_{\! #1}}
\newcommand{\ZN}{Z_N}
\newcommand{\uGCE}{u_{\mathrm{GCE}}}
\renewcommand{\d}{\mathrm d}

\begin{document}

\begin{frontmatter}

\title{Quantum canonical ensemble: a projection operator approach}

\author[adres1,adres2]{Wim Magnus\corref{correspondingauthor}}
\cortext[correspondingauthor]{Corresponding author}
\ead{wim.magnus@uantwerpen.be}

\author[adres3]{Lucien Lemmens}
\ead{lcnlmmns@me.com}

\author[adres3]{Fons Brosens}
\ead{fons.brosens@uantwerpen.be}

\address[adres1]{imec, Kapeldreef 75, B-3001 Leuven, Belgium}
\address[adres2]{Universiteit Antwerpen, Physics Department,
                 Groenenborgerlaan 171, B-2020 Antwerpen, Belgium}
\address[adres3]{Universiteit Antwerpen, Physics Department,
                 Universiteitsplein 1, B-2060 Antwerpen, Belgium}

\begin{abstract}
Fixing the number of particles $N$, the quantum canonical ensemble imposes
a constraint on the occupation numbers of single-particle states. The
constraint particularly hampers the systematic calculation of the partition
function and any relevant thermodynamic expectation value for arbitrary $N$
since, unlike the case of the grand-canonical ensemble, traces in the
$N$-particle Hilbert space fail to factorize into simple traces over
single-particle states.
In this paper we introduce a projection operator that enables a
constraint-free computation of the partition function and its derived
quantities, at the price of an angular or contour integration.
Being applicable to both bosonic and fermionic systems in
arbitrary dimensions, the projection operator approach provides transparent
integral representations for the partition function $Z_N$ and the Helmholtz
free energy $\FH{N}$ as well as for two- and four-point correlation
functions. While appearing only as a secondary quantity in the present
context, the chemical potential emerges as a by-product from the relation
$\muup_N = \FH{N+1} - \FH{N}$, as illustrated for a two-dimensional
fermion gas with $N$ ranging between 2 and 500.
\end{abstract}


\end{frontmatter}


\section{Introduction}
The calculation of the quantum mechanical partition function $\ZN$ of $N$
identical particles treated in the framework of the canonical ensemble
remains a long-standing problem in many-body theory, even if the particles
do not interact. The main difficulty hampering a systematic evaluation of
$\ZN$ for moderate to large values of $N$ originates from the particle
number constraint that is to be invoked explicitly. In order to overcome
this problem, we introduce a projection operator in
section~\ref{sec:Z_canonical} which is capable of dealing with the particle
number constraint for non-interacting particles (bosons, fermions) as well
as systems of interacting particles complying with particle number
conservation.
However, the formal applicability to interacting particles is hardly
useful in practice, because the eigenstates and the eigenvalues of the
energy for such systems are rarely available.
Although modern particle physics surely treats strongly interacting
particles, it faces the necessity of applying approximations which,
in essence, apply a variety of transformation techniques that reduce the
problem to treating ensembles of non-interacting particles.
Thermal expectation values, based on statistical averages over ensembles of
non-interacting particles, still provide the generic building blocks to set
up perturbational and variational as well as other non-perturbative
computation schemes.
Essential ingredients for such approaches are the partition function and
the two- and four-point correlation functions characterizing systems of
non-interacting particles.

Keeping all this in mind, we believe it remains utterly relevant
to consider a system of non-interacting particles and, therefore, we first
examine its canonical partition function.
As detailed in section~\ref{sec:Independent}, this results into a
transparent integral representation for the partition function as well as
the corresponding Helmholtz free energy and, hence, the chemical potential
of $N$ non-interacting fermions or bosons.
The integral representation also allows for a very simple derivation
of a known recurrence relation for the partition function.

For $N$ harmonic oscillators in 1 dimension, the partition function could
be obtained in closed form.
The results are presented in section~\ref{sec:HO1D}, for bosons as well as
for fermions.

Section~\ref{sec:2Delgas} contains a few numerical results related to the
partition function and derived quantities of a finite size two-dimensional
electron gas.
Finally, the projection operator approach is applied once more in
section~\ref{sec:corrfunc} to derive generic expressions for the two- and
four-point correlation functions. 
Some rather technical aspects are redirected to two appendices.

\section{The canonical partition function: a projection operator approach
         \label{sec:Z_canonical}}
According to the nomenclature developed in the beginning of the 20th
century, the statistical knowledge of a system in thermal equilibrium
depends on the ensemble type: microcanonical, canonical or grand canonical.
The canonical ensemble assumes that the exact number of particles in the
system is known while its grand canonical counterpart merely requires that
the average particle number be available. In theoretical studies of nuclear
systems the number of particles is intrinsically dictated by the problem
while for a great majority of solid-state systems only the average number
of particles, in casu the density, is relevant. \\
However, recent technological developments in nanoelectronics made it
possible to control the number of carriers in nanometer-scaled devices,
making the actual number of particles a more important parameter than the
average number or density. Hence it would be desirable to export and extend
theoretical methods developed in nuclear physics to various many-body
formalisms commonly used to treat nanometer-scaled solids.
A typical many body approach often starts with a short investigation of the
non-interacting system, usually formulated in terms of creation and
annihilation operators. The use of these operators implicitly invokes a
Fock space that, by construction, discards any reference to the number of
particles whatsoever.
However, if a description with a fixed number is mandatory, one needs to
introduce a projection technique that limits the Fock space to a subspace
that corresponds to a fixed number of particles, while still allowing for a
formulation in terms of the second quantization operators.
The projection technique used for nuclear models can accomplish this task
and is found to operate also for the second quantization description of a
many-body Hamiltonian. Correspondingly, the number of particles is fixed
and emerges as a fixed eigenvalue of the number operator.
After the projection one has to focus on the Fock subspace that is
exclusively related to a fixed number of particles. In particular, the
many-particle eigenfunctions of the projected Hamiltonian have to be
calculated together with their energy spectrum and, afterwards, the
probability density.

Motivated by the above observations, we consider a fixed number $N$ of
of indistinguishable particles, fermions or bosons, described in the
many-particle Fock space by a second-quantized Hamiltonian $\Hop$. 
In order to preserve the number of particles, $\Hop$ has to commute with
the particle number operator $\Nop$. Consequently, due to
$[\Hop, \Nop] = 0$, many-particle eigenstates $\fket{\psi_{j, N}}$ of
$\Hop$ can be found that simultaneously diagonalize $\Hop$ and $\Nop$,
i.e.
\begin{equation}
   \Hop \fket{\psi_{j, N}} = \EjN \fket{\psi_{j, N}}, \qquad
   \Nop \fket{\psi_{j, N}} =    N \fket{\psi_{j, N}}, \qquad
   N = 0, 1, 2, 3, \ldots
   \label{eq:Z_canonical:H&Noperators}
\end{equation}
Representing an arbitrary, allowable number of particles, the eigenvalues
$N$ of $\Nop$ are used to label the eigenstates $\fket{\psi_{j, N}}$ as
well as the corresponding $\EjN$. The index $j$ co\-vers all remaining,
internal quantum numbers that are labeling $\EjN$ for a fixed value of $N$.
For the sake of notational simplicity, we have omitted below any dependence
on spin components which, however, can be incorporated into the formalism
whenever required.
Because $\Hop$ operates in Fock space without any a priory reference to
the number of particles, thermodynamics is usually expressed in the grand
canonical ensemble (GCE). Within this framework, the chemical potential
emerges as a Lagrange multiplier regulating the average number of
particles, rather than imposing a sharply defined value of $N$, as required
in the canonical ensemble (CE).
In order to overcome this problem, we propose a projection operator that
extracts a $N-$particle Hamiltonian $\HNpart$ out of $\Hop$, while
automatically invoking the canonical constraint of $N$ particles.

Let $\left\{ \ket{\psi_{n, M}} \right\}$ denote the complete set of
eigenstates with an integer, nonnegative eigenvalue $M$ of the number
operator $\Nop\mathrm{:}$
\begin{equation}
   \Nop \ket{\psi_{n, M}} = M \ket{\psi_{n, M}}, \quad \forall n,
\end{equation}
and consider the operator
\begin{equation}
   \PN = \frac{1}{2 \pi} \int_{-\pi}^{\pi} \!\!
         \exp \left( \im (\Nop - N) \theta \right) \d \theta,
   \label{eq:Z_canonical:PN_def}
\end{equation}
with the obvious properties $\hat{\mathsf{P}}_{N}^{\dagger} = \PN$
and $\hat{\mathsf{P}}_{N}^2 = \PN$. One furthermore observes that
\begin{equation}
   \PN \ket{\psi_{n, M}} = \frac{1}{2 \pi}\int_{-\pi}^{\pi} \!\!
   \exp \left( \im (M - N) \theta \right) \d \theta \ket{\psi_{n, M}} =
   \delta_{N, M} \ket{\psi_{n, N}}
\end{equation}
and hence
\begin{equation}
   \PN \sum_{n, M} A_{n, M} \ket{\psi_{n, M}} = A_{n, N} \ket{\psi_{n, N}}.
\end{equation}
Therefore, $\PN$ is a real projection operator which yields an eigenstate of
the $N$-particle subspace if it acts on an arbitrary state of the entire
many-particle Hilbert space. Consequently
\begin{equation}
   \HNpart = \PN \Hop \, \PN
\end{equation}
is the $N$-particle Hamiltonian, extracted from the Hamiltonian $\Hop$
in the many-particle Hilbert space.
Note that the position representation of $\HNpart$ in principle coincides
with the $N$-particle Hamiltonian of first quantization, as can be inferred
from the algebraic treatment given in Ref.~\cite{Robertson}.
Although similar projection operators have been introduced before in
statistical physics~\cite{ElzeGreiner1}, nuclear and high-energy
physics~\cite{ElzeGreiner2, Elzeetal, Benderetal}, we are not
aware of its practical use as a particle number regulator in quantum
statistics.

It is tempting to immediately suppose that the partition function for
thermodynamical equilibrium is given by
\begin{equation}
   \ZN(\beta) = \Sp \left( \e^{-\beta \HNpart} \right)
\end{equation}
with some typical derived quantities as the Helmholtz free energy
$\FN(\beta)$ and the internal energy $\UN(\beta)$, with
$\beta = 1 / (\kB T)$, where is the Boltzmann constant
$(\kB = 1.3806568 \times 10^{-23} \unit{J} \unit{K}^{-1})$ and $T$ the
temperature in Kelvin:
\begin{align}
   \FN(\beta)
   & = -\frac{1}{\beta} \ln \left( \ZN(\beta) \right)
       \label{eq:Z_canonical:FNfromZN} \\
   \UN(\beta)
   & = -\frac{\d}{\d \beta} \ln \left( \ZN(\beta) \right).
       \label{eq:Z_canonical:UNfromZN}
\end{align}
Although correct, these equation should be handled with care. Thermal
equilibrium means that the internal energy $\UN$ is stable in time, and
$\beta$ (and hence $T$) is in essence a Lagrange multiplier for imposing
that condition, rather than a given quantity. The internal energy $\UN$
is the fixed quantity. Because of the technicality of this question, the
correct interpretation of the principle of maximum entropy
\cite{Jaynes1,Jaynes2,Grandy} in thermal equilibrium is treated
in~\ref{sec:app:Entropy}.

This approach is clearly consistent with~(\ref{eq:Entropy:ZN(UN)}) under the
condition~(\ref{eq:Entropy:UNwith_beta(UN)}). Defining the generating
function
\begin{equation}
   G(\beta, \theta) =
   \Sp \left( \e^{-\beta \Hop} \e^{\im \Nop \theta} \right),
   \label{eq:Z_canonical:G_def}
\end{equation}
and comparing it with the definition~(\ref{eq:Z_canonical:PN_def}) of
$\PN$, we obtain $\ZN(\beta)$ as the $N$-th coefficient of the Fourier
series that represents $G(\beta, \theta)$,
\begin{equation}
   \ZN(\beta) =
   \frac{1}{2 \pi} \int_{-\pi}^{\pi} \, G(\beta, \theta) \e^{-\im N \theta}
   \d \theta.
   \label{eq:Z_canonical:ZNinG}
\end{equation}

\section{Partition function of non-interacting indistinguishable particles
\label{sec:Independent}}

As already argued in the Introduction, the general formulation of the
previous section, though valid also for interacting particles, is of
limited practical use. Quantum statistics of non-interacting particles on
the other hand, still provides the basic ingredients for most approximative
treatments of interacting particles. Therefore, we first concentrate on the
partition function of non-interacting particles with supposedly known
eigenstates and energy levels. The Hamiltonian $\Hop$ and the number
operator $\Nop$ can then be expressed in terms of the single-particle
energy spectrum $\epsk$, where $k$ denotes any set of generic quantum
numbers properly labeling the single-particle energies and the
corresponding eigenfunctions:
\begin{equation}
   \Hop = \sum_k \hat{n}_{k}^{\vphantom{\dagger}} \epsk, \qquad
   \Nop = \sum_k \hat{n}_{k}^{\vphantom{\dagger}}, \qquad
   \hat{n}_{k} = \ckd \ck
   \label{eq:Independent:H&Nin_nk}
\end{equation}
where the creation and destruction operators $\ckd$ and $\ck$ satisfy
appropriate (anti)commu\-tation relations, i.e.
\begin{alignat}{3}
   \left[ \ckd, \ckpd \right]
   & = 0 = \left[ \ck, \ckp \right], \qquad
   & \left[ \ck, \ckpd \right] = \delta_{k, k^{\prime}} \quad
   & \text{for bosons,} \\
   \left\{ \ckd, \ckpd \right\}
   & = 0 = \left\{ \ck, \ckp \right\}, \qquad
   & \left\{ \ck, \ckpd \right\} = \delta_{k, k^{\prime}} \quad
   & \text{for fermions.}
\end{alignat}
This means that any particular energy $\EjN$
in~(\ref{eq:Z_canonical:H&Noperators}) takes the form
\begin{equation}
   \EjN = \sum_k \nk \epsk \qquad \text{with} \qquad \sum_k \nk = N,
\end{equation}
the integer occupation numbers $\nk$ being restricted to 0 and 1 for
fermions while ranging between 0 and infinity for bosons. Keeping the total
number of particles fixed is prohibitive~\cite{Feynman} for writing $Z_{N}$
as $\prod_k \sum_{\nk} \exp(-\beta \nk \epsk)$. As can be found in many
textbooks, e.g., in Ref.~\cite{Kleinert}, the standard approach to remedy
this problem involves the construction of all cyclic decompositions of the
particle permutations, which turns out to be a tedious task. Use of the
projection operator greatly simplifies this conditional summation. Elementary
operator algebra enables one to work out~(\ref{eq:Z_canonical:G_def})
explicitly, yielding
\newline
$G(\beta, \theta) =
\Sp \left( \prod_k \e^{(\im \theta - \beta \epsk) \, \nkop} \right) =
\prod_k \Sp \left( \e^{(\im \theta - \beta \epsk) \, \nkop} \right)$, and thus
\begin{equation}
   G(\beta, \theta) = \prod_k \sum_{\nk}
                      \exp \left( (\im \, \theta - \beta \epsk) \nk \right).
\end{equation}
Summing $\nk$ from 0 to $\infty $ for bosons, and from 0 to 1 for fermions,
readily gives
\begin{alignat}{3}
   G(\beta, \theta)
   & = \prod_k
       \left( 1 - \exp \left( \im \theta - \beta \epsk \right) \right)^{-1}
       \quad
   & & \text{for bosons,} \\
   G(\beta, \theta)
   & = \prod_k
       \left( 1 + \exp \left( \im \theta - \beta \epsk \right) \right)
       \quad
   & & \text{for fermions,}
\end{alignat}
which (less transparent but more compact) can be abbreviated as
\begin{equation}
   G(\beta, \theta) = \prod_k
   \left( 1 - \xi \exp \left( \im \theta - \beta \epsk \right) \right)^{-\xi}
   \text{with}
   \begin{cases}
      \, \xi = +1 & \text{for bosons,} \\
      \, \xi = -1 & \text{for fermions.}
   \end{cases}
   \label{eq:Independent:G}
\end{equation}
Filling this out in~(\ref{eq:Z_canonical:ZNinG}), it should be noted that
the angular integral can equivalently be expressed as a complex contour
integral along the unit circle
\begin{equation}
    \ZN(\beta) =
    \frac{1}{2 \pi \im} \oint_{\left \vert z \right \vert = 1}
    \frac{\tilde{G}(\beta, z)}{z^{N+1}} \, \d z, \qquad
    \tilde{G}(\beta, z) =
    \prod_k \left( 1 - \xi z \, \e^{-\beta \epsk} \right)^{-\xi}.
    \label{eq:Independent:ZinGtilde}
\end{equation}
The generating function $\tilde{G}(\beta, z)$ is analytic everywhere for
fermions ($\xi =-1$) whereas, for bosons, the region $|\, z \, | \leqslant 1$
would merely contain an isolated singularity at $z = 1$ if the single-particle
ground-state energy were vanishingly small. In order not to introduce spurious
poles, all boson single-particle eigenenergies should be strictly positive.
This can always be realized by an energy shift resulting from a gauge
transformation. This ensures that, inside the unit circle, the integrand of
Eq.~(\ref{eq:Independent:ZinGtilde}) has a single pole at $z = 0$, whence
\begin{equation}
   \ZN(\beta) = \frac{1}{N!} \lim_{z\rightarrow 0}
                \frac{\partial^N \tilde{G}(\beta, z)}{\partial z^N}.
   \label{eq:ZNpole}
\end{equation}
Obtaining first the derivative of $\ln \tilde{G}(\beta ,z)$ to get
\begin{equation}
   \frac{\partial \tilde{G}(\beta, z)}{\partial z} =
   \tilde{G}(\beta, z) \sum_k \frac{1}{\e^{\, \beta \epsk}-z \xi},
   \label{eq:Independent:derivGtilde}
\end{equation}
we apply Leibniz' rule to take the $(N-1)$-th derivative of
Eq~(\ref{eq:Independent:derivGtilde}) for $N \geqslant 1$ to arrive at
\begin{multline}
   Z_0(\beta) = 1, \quad Z_1(\beta) = \sum_k \exp(-\beta \epsk),
   \label{eq:Independent:ZNrecurrence} \\
   \ZN(\beta) = \frac{1}{N} \sum_{j = 0}^{N - 1} \xi^{N - j - 1}
   Z_{\! j}(\beta) \, Z_1((N - j) \beta) \quad \text{for $N \geqslant 1$}.
\end{multline}
This recurrence relation is not new \cite{Landsberg}~\cite{BormannFranke},
but its derivation from the contour integral~(\ref{eq:Independent:ZinGtilde})
is substantially simpler than what follows from a tedious analysis of the
permutation group. It also enhances the confidence in the correctness of
the projection operator approach.

Given the occurrence of the variable $z$ as a prefactor of the exponentials
$\e^{-\beta \epsk}$ in $\tilde{G}(\beta, z)$, it might be tempting to
interpret $z$ as a complex fugacity in analogy with the real fugacity
$\exp(\beta \muGCE)$ appearing similarly in the grand-canonical partition
function and the Bose-Einstein and Fermi-Dirac distribution functions.
However, a safer interpretation could lie in the comparison of the CE and
the GCE: whereas the latter sets the chemical potential to fix the average
number of particles arising consequently as a weighted sum over all particle
numbers $N$, the CE fixes $N$ and is therefore bound to integrate over all
relevant \textquotedblleft complex fugacities\textquotedblright.

To clarify this point, we extend the unit circle in
(\ref{eq:Independent:ZinGtilde}) to another circular contour with radius
$u > 0$:
\begin{align}
   \ZN(\beta)
   & = \frac{1}{u^{N}} \frac{1}{2 \pi} \int_{-\pi}^{\pi}
       \tilde{G}(\beta, u \, \e^{\im \theta}) \, \e^{-\im \theta N} \,
       \d \theta
       \nonumber \\
   & = \frac{\tilde{G}(\beta, u)}{u^{N}} \; \times \;
       \frac{1}{2 \pi} \int_{-\pi}^{\pi}
       \frac{\tilde{G}(\beta, u \e^{\im \theta})}{\tilde{G}(\beta, u)}
       \e^{-\im \theta N} \, \d \theta.
       \label{eq:Independent:ZNz2u}
\end{align}

The fact that this expression is independent of $u$ implies
$\partial Z_N(\beta) / \partial u = 0$.
Because of Eq.~(\ref{eq:Independent:derivGtilde}), this means that
\begin{equation}
   \int_{-\pi}^{\pi}
   \tilde{G}(\beta, u \e^{\im\theta}) e^{-\im N \theta}
   \left(
         N - \sum_k 
         \left(
               \frac{1}{u} \e^{\beta \epsk - \im \theta}
              -\xi
         \right)^{-1}
   \right)
   \d \theta = 0.
\end{equation}
The above sum rule for the CE can not be satisfied by $\uGCE$, the value of
$u = \exp(\beta \muup)$ that solves the transcendental equation for the
GCE, i.e.
\begin{equation}
   N = \sum_k \,
       \left( \e^{-\beta \muup} \e^{\beta \epsk} -\xi \right)^{-1}.
       \label{eq:Independent:N,mu_transcendental}
\end{equation}
Consequently, in the light of the CE,
Eq.~(\ref{eq:Independent:N,mu_transcendental}) should be considered an
approximative equation, usually obtained from a saddle point method. The
latter amounts to maximizing the factor $\tilde{G}(\beta, u) / u^N$ in the
second line of Eq.~(\ref{eq:Independent:ZNz2u}), where it is expected that
$Z_N(\beta) \approx \tilde{G}(\beta, u_{GCE}) / u_{GCE}^{N}$ gives a good
estimate of the free energy. And indeed, the Helmholtz free energy then
becomes
\begin{equation}
   \FN(\beta) = 
   \begin{dcases}
      & \frac{1}{\beta}
      \sum_k \ln \left( 1 - \e^{\, \beta(\muup - \epsk)} \right) + N \muup
      \quad \text{for bosons,} \\[4 mm]
      \, - \hskip -3 mm
      & \frac{1}{\beta}
      \sum_k \ln \left( 1 + \e^{\, \beta(\muup - \epsk)} \right) + N \muup
      \quad \text{for fermions.}
   \end{dcases}
\end{equation}
For sufficiently large $N$ this is consistent with the familiar assumption
$\muup \approx F_{N + 1}(\beta) - \FN(\beta)$, since then
$\muup_{N+1} \approx \muup_N$.
But the present derivation clearly shows how and why the standard
transition from the CGE to the CE is an approximation. A correct treatment
of the CE has to deal with the angular integral or, equivalently, the
complex contour integral for $\ZN$ (or its equivalent recurrence
relations).

\section{Indistinguishable harmonic oscillators in 1D\label{sec:HO1D}}

Until now, closed form solutions involving indistinguishable particles
are barely available, even if they are not interacting. As an exception,
however, we illustrate the case of non-interacting bosons and fermions
collectively moving in a 1D harmonic potential and sharing the well-known
single-particle energy spectrum
\begin{equation}
   \epsk = \left( k + \frac{1}{2} \right) \hbar \omega, \quad
           k = 0, 1, \ldots, \infty.
\end{equation}
The countour integral representation (\ref{eq:Independent:ZinGtilde}) or,
equivalently, the derivative rule (\ref{eq:ZNpole}) relates $\ZN(\beta)$ to
the generating function $\Gt(\beta, z)$. In the present case, the latter is
given by
\begin{equation}
   \Gt(\beta, z) =
   \begin{cases}
      \displaystyle
      \, \prod_{k = 0}^{\infty} \frac{1}{1 - z \, \e^{-\beta \epsk}} =
         \prod_{k = 0}^{\infty}
         \frac{1}{1 - z \, \e^{-\beta \hbar \omega (k + 1/2)}}
      & \quad \text{for bosons,} \\[4 mm]
      \displaystyle
      \, \prod_{k = 0}^{\infty} \left( 1 + z \, \e^{-\beta \epsk} \right) =
         \prod_{k = 0}^{\infty}
         \left( 1 + z \, \e^{-\beta \hbar \omega (k + 1/2)} \right)
      & \quad \text{for fermions.}
   \end{cases}
\end{equation}
Direct evaluation of the $N$-th derivative of $\Gt(\beta, z)$ seems quite
a formidable task, if possible at all. However, two mathematical
identities derived by Leonhard Euler and nowadays emerging as corrolaries
of the $q$-binomial theorem \cite{Andrews,Bellman} (see also
\ref{sec:app:Euler}) are found to solve the problem. According to the
identity (\ref{eq:euler1}), the infinite product for bosons can be written
as a convergent series for $|z| < 1$. Similarly, the identity
(\ref{eq:euler2}) can be used for the fermionic case. The result is
\begin{equation}
   \Gt(\beta, z) =
   \begin{cases}
      \displaystyle
      1 + \sum_{n = 1}^{\infty}
      \left( z \e^{-\beta \hbar \omega / 2} \right)^n
      \prod_{k = 1}^n \frac{1}{1 - \e^{-\beta \hbar \omega k}}
      & \quad \text{for bosons,} \\[4 mm]
      \displaystyle
      1 + \sum_{n = 1}^{\infty}
      \left( z \e^{-\beta \hbar \omega / 2} \right)^n
      \e^{-\beta \hbar \omega n (n - 1) /2}
      \prod_{k = 1}^n \frac{1}{1 - \e^{-\beta \hbar \omega k}}
      & \quad \text{for fermions.}
   \end{cases}
\end{equation}
In accordance with Eq.~(\ref{eq:ZNpole}) the coefficient of $z^N$ in the
above series is the partition function for $N$ oscillators:
\begin{equation}
   \ZN(\beta) =
   \begin{cases}
      \displaystyle
      \e^{-N \beta \hbar \omega / 2} \prod_{k = 1}^N
      \frac{1}{1 - \e^{-\beta \hbar \omega k}}
      & \quad \text{for bosons,} \\[4 mm]
      \displaystyle
      \e^{-N^2 \beta \hbar \omega / 2} \prod_{k = 0}^{\infty}
      \left( 1 - \e^{-\beta \hbar \omega k} \right)
      & \quad \text{for fermions.}
   \end{cases}
\end{equation}
Having determined the partition function, one may easily find the
Helmholtz free energy
\begin{equation}
   \FN(\beta) = \frac{1}{\beta} \sum_{k = 1}^N
                \ln \left( 1 - \e^{-\beta \hbar \omega k} \right) +
   \begin{dcases}
      \frac{1}{2} N   \hbar \omega & \text{for bosons,} \\
      \frac{1}{2} N^2 \hbar \omega & \text{for fermions.}
   \end{dcases}
\end{equation}
Complying with the standard definition $\muN = F_{N + 1} - \FN$ of the
chemical potential, one thus readily obtains
\begin{equation}
   \muN(\beta) = \frac{\ln \left( 1 - \e^{-(N + 1) \beta \hbar \omega} \right)}  {\beta}
+   \begin{dcases}
        \, \hbar \omega
        & \text{for bosons,} \\
        \, \left( N + \frac{1}{2} \right) \hbar \omega
        & \text{for fermions,}
    \end{dcases}
\end{equation}
clearly depending on both $N$ and $\beta$. Only for sufficiently large $N$, more
precisely for $\e^{-N\beta \hbar \omega} \ll1$, the logarithmic term can be
neglected, such that
\begin{equation}
   \muN(\beta) \underset{\e^{-N\beta \hbar \omega} \ll 1}{\approx }
   \begin{dcases}
      \, \hbar \omega
      & \text{for bosons,} \\
      \, \left( N + \frac{1}{2} \right) \hbar \omega
      & \text{for fermions.}
   \end{dcases}
\end{equation}
For the internal energy, one finds
\begin{equation}
   \UN(\beta) = \sum_{k = 1}^N
                \frac{k \hbar \omega }{\e^{\beta \hbar \omega k} - 1} +
   \begin{dcases}
      \frac{1}{2} N   \hbar \omega & \text{for bosons,} \\
      \frac{1}{2} N^2 \hbar \omega & \text{for fermions.}
   \end{dcases}
\end{equation}

As discussed in~\ref{sec:app:Entropy}, this equation should be considered
as a transcendental equation, determining $\beta$ for given $\UN$.
Nevertheless, it is common practice to express $\UN$ as a function of
$\beta$, in which case it would just take a simple rotation of the
corresponding $\UN(\beta)$ curve to obtain the requested $\beta(\UN)$
relation. \\
However, it is more instructive to look at the specific heat
$\d \UN / \d T$, using $\beta = 1 / \left( \kB T \right)$:
\begin{equation}
   \frac{\d \UN}{\d T} = \kB \sum_{k = 1}^N
   \frac{\left( k / \tau \right)^2 \e^{k / \tau}}
        {\left( \e^{k / \tau } - 1 \right)^2}
   \quad \text{with} \quad \tau = \frac{\kB T}{\hbar \omega},
\end{equation}
which holds for both fermions and bosons. It is clear that
$\d \UN / \d T \underset{T \rightarrow \infty}{\rightarrow} N \kB$
as expected. The important point to be emphasized is that the relation
between $\UN$ and the temperature is in general \textsc{not} linear. The
convergence to the classical limit $\UN \rightarrow N \kB T$ even slows
down with increasing $N$.

\section{A two-dimensional electron gas \label{sec:2Delgas}}

The formal expressions for the partition function obtained in sections
\ref{sec:Independent} clearly show that any practical investigation of
statistical physics within the framework of the canonical ensemble is
bound to deal with the angular integral or, equivalently, the complex
contour integral. It occurs in the expression of all thermodynamical
quantities (partition function, free energy, specific heat, \ldots), either
by direct evaluation or by conversion into the equivalent recurrence
relations. Analytical results can only be expected for an extremely small
number of systems. The previous section gave such an example, but in
general one has to rely on numerical methods.

As an illustration we quote the calculation of the free energy and the
chemical potential for a free electron gas residing in a finite,
two-dimensional rectangular area
$0 \leqslant x \leqslant L_x, \; 0 \leqslant y \leqslant L_y$,
while imposing periodic boundary conditions on the single-electron wave
functions and taking the single-electron energy spectrum to be
\begin{equation}
   \epsk \to \epsilon_{n_x, n_y}
   = \frac{\hbar^2}{2 m}
     \left[
           \left( \frac{2 \pi n_x}{L_x} \right)^2 +
           \left( \frac{2 \pi n_y}{L_y} \right)^2
     \right],
     \quad n_x, n_y = 0, \pm 1, \pm 2, \ldots
\end{equation}
with $m$ the electron mass.

In view of possible practical applications, for example for the electron
gas in the inversion layer of a MOS field-effect transistor, we have fixed
$L_x$ and $L_y$ to be 100 nm, whereas the ambient temperature is assumed to
be 300~K.
\begin{figure}
   \includegraphics[width=0.95\textwidth]{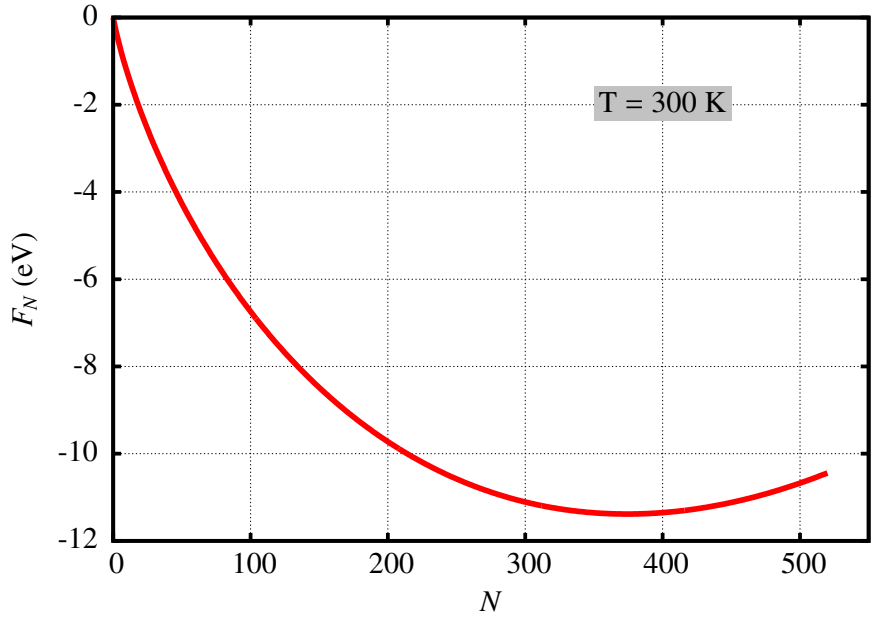}
   \caption{Helmholtz free energy for $L_x = L_y = 100$ nm.}
   \label{fig:F520}
\end{figure}
Extracted from the recurrence relation (\ref{eq:Independent:ZNrecurrence}),
the Helmholtz free energy is plotted in Fig.~(\ref{fig:F520}) versus the
number of electrons $N$, up to $N = 520$ beyond which sign changes (due to
$\xi = -1$ for fermions) made the recurrence relation unstable. It turns
out that, for fixed $L_x = L_y$, the free energy attains a minimum for a
particular value of $N$ -- in the present case around $N = 374$ -- which
corresponds to $\muup_N = 0$ or, equivalently, the absence of energy cost
when a single particle is to be added or removed. On the other hand, a
typical value of the areal electron concentration in a MOS capacitor
operating at room temperature is $10^{12}$ cm$^{-2}$ which, for
$L_x = L_y = 100$ nm, corresponds to $N \approx 100$ and, hence, to a
negative chemical potential.

\begin{figure}
   \includegraphics[width=0.95\textwidth]{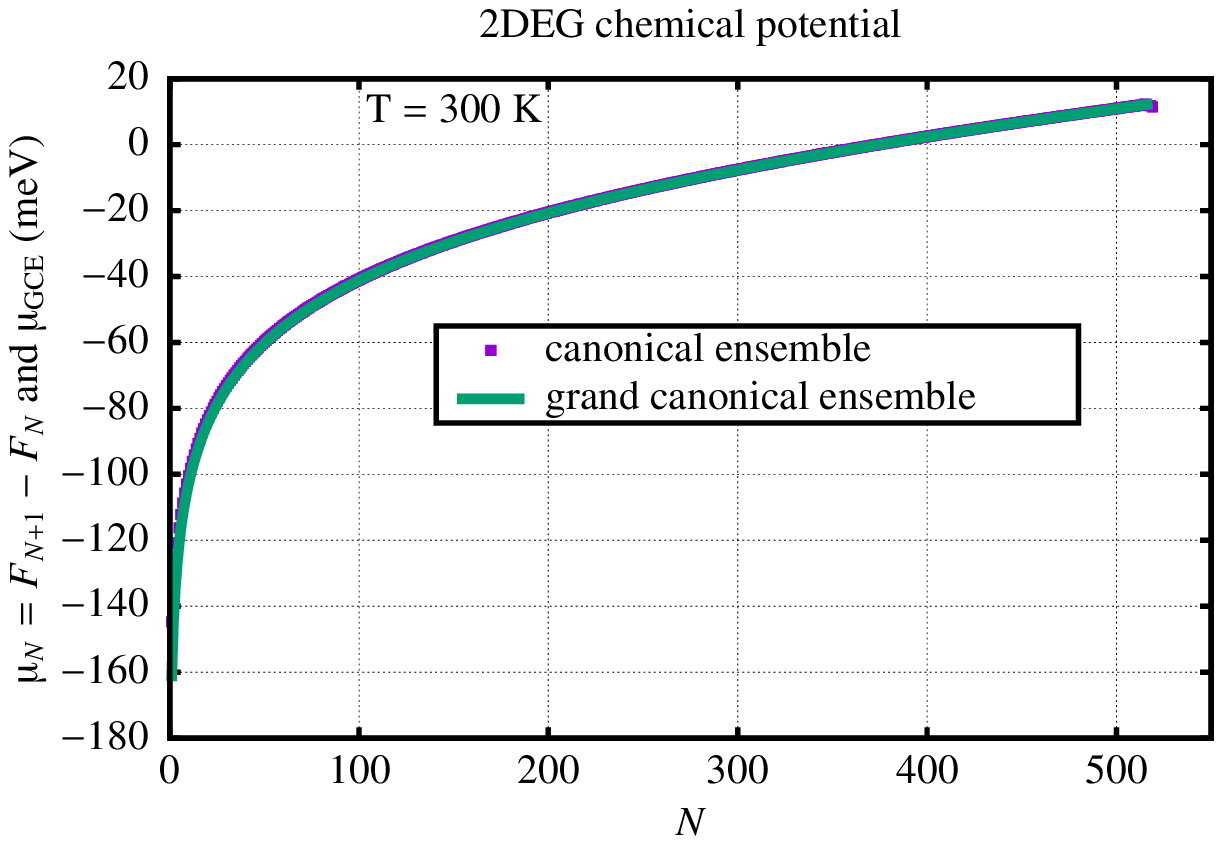}
   \caption{Canonical and grand-canonical chemical potentials for
            $L_x = L_y = 100$ nm up to $N = 520$.}
   \label{fig:mu520}
\end{figure}

\begin{figure}
   \includegraphics[width=0.95\textwidth]{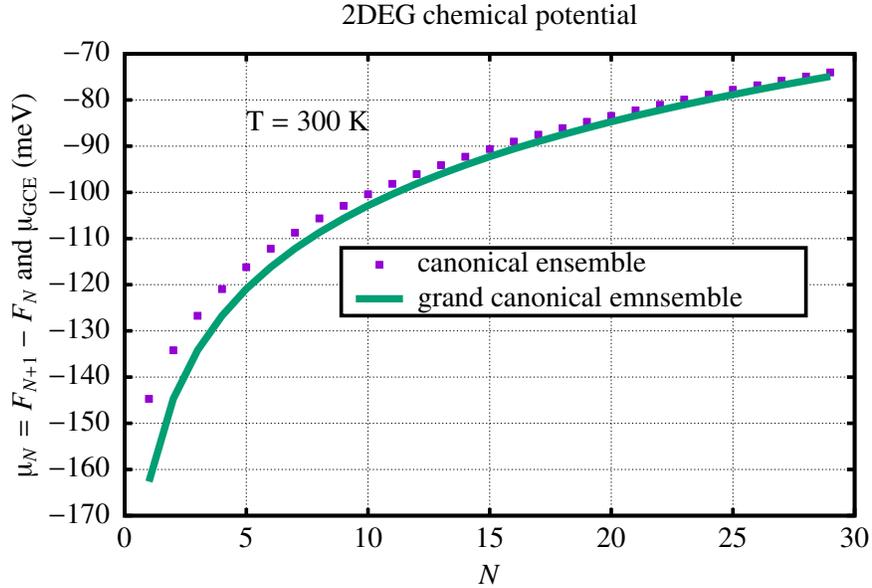}
   \caption{Canonical and grand-canonical chemical potentials for
            $L_x = L_y = 100$ nm up to $N = 30$.}
    \label{fig:mu30}
\end{figure}

Unlike $\muup_N$, the grand-canonical chemical potential $\muGCE$ that
corresponds to the thermodynamic limit $N \to \infty, L_x, L_y \to \infty$,
whilst $\ns \equiv N / (L_x L_y)$ remains finite, can be calculated
analytically from
\begin{equation}
   \muGCE = \frac{1}{\beta} \ln
            \left(
                  \exp \left( \frac{2 \pi \beta \hbar^2 \ns} {m} \right)
                  - 1
            \right).
   \label{eq:muGCE}
\end{equation}
For the sake of comparison we have plotted both $\muup_N$ and $\muGCE$
versus $N$ in Fig.~\ref{fig:mu520} and in Fig.~\ref{fig:mu30}.
For relatively large values of $N$, say $N > 30$, $\muup_N$ and $\muGCE$
are about equal. On the other hand, the expression (\ref{eq:muGCE}) for
$\muGCE$ is only valid in the thermodynamic limit.
It fails to characterize electron ensembles with $N \leqslant 30$ which
can, however, be handled by the canonical formalism yielding $\muup_N$.

Moreover, in the case of more complicated fermionic systems, such as the
3DEG, the transcendental
equation~(\ref{eq:Independent:N,mu_transcendental}) expressing the
(average) number of fermions in terms of $\muGCE$ generally can no longer
be inverted analytically, while the computational scheme that yields
$\muup_N$ remains unaltered.

\section{Correlation functions \label{sec:corrfunc}}

So far we have concentrated on the projection operator approach for
obtaining the partition function and derived quantities of
indistinguishable particles, with particular attention to non-interacting
particles. Two specific examples were worked out. But, as already mentioned
in the Introduction, not just the partition function and its derived
quantities but also single- and two-particle correlation functions of
non-interacting particles are instrumental to perturbative and variational
methods that are commonly entering approximative treatments of interacting
particles. In the present section, it is shown that the projection operator
is also well equipped to calculate these quantities.

Single- and two-particle correlation functions -- also referred to as
two-point and four-point functions -- typically provide a signature of the
correlation between particles that are spatially separated. For the sake of
notational simplicity, positions in space are denoted by $x$ and
$x^{\prime }$ which, however, should not at all be regarded as a limitation
to strictly one-dimensional systems. \newline
Considered quantum statistical averages, correlation functions are
conveniently expressed in terms of field operators $\psi(x)$ satisfying
typical (anti)commutation relations
$\{\psi(x), \psi^{\dagger}(x^{\prime})\} = \delta(x - x^{\prime })$ etc.

\subsection{Single-particle correlation functions}

Adopting once again the canonical ensemble framework, we can reinvoke the
above defined projection operator $\PN$ to calculate the single-particle
correlation function (two-point function) for an ensemble of $N$ particles
from
\begin{equation}
   \Stwo(x, x')
   = \fbraket{\psid(x) \, \psi(x')}
   \equiv \frac{1}{Z_N}
   \Sp \left( \PN \exp (-\beta \, \Hop) \, \psid(x) \, \psi(x') \right).
   \label{eq:Stwo}
\end{equation}
In most cases of interest, both the single-particle energies $\epsk$ and the
corresponding single-particle wave functions $\phik(x)$ are supposed to be
explicitly known, the latter constituting a complete, orthonormal basis.
Hence, it proves convenient to expand the field operators as
\begin{equation}
   \psi(x') = \sum_{k'} \phikp(x') \, \ckp, \quad
   \psid(x) = \sum_k    \phikc(x)  \, \ckd,
\end{equation}
the creation and destruction operators thereby appearing as expansion
coefficients. Substitution into (\ref{eq:Stwo}) yields
\begin{equation}
   \Stwo(x, x') = \sum_{k k'} \phi^*_k(x) \, \phi_{k'}(x') \fbraket{\ckd \ckp}
   \label{eq:Stwoxxp}
\end{equation}
with
\begin{align}
   \fbraket{\ckd \ckp}
   & = \frac{1}{Z_N}
       \Sp
       \left( \PN \exp \left( -\beta \, \Hop \right) \ckd \ckp \right)
       \nonumber \\
   & = \frac{1}{2 \pi Z_N}
       \int_{-\pi}^{\pi} \!\! \d \theta \,
       \exp \left( -\im N \theta \right)
       \Sp
       \left(
             \exp \left( \im \theta \Nop -\beta \, \Hop \right)
             \ckd \ckp
       \right)
       \nonumber \\
   & = \frac{1}{2 \pi Z_N}
       \int_{-\pi}^{\pi} \!\! \d \theta \,
       \exp \left( -\im N \theta \right) \Lambda_{k k'}(\theta).
\end{align}
In order to evaluate the trace
\begin{equation}
   \Lambda_{k k'}(\theta)
   = \Sp
     \left(
           \exp \left( \im \theta \, \Nop -\beta \, \Hop \right) \ckd \ckp
     \right),
\end{equation}
we first exploit its invariance under cyclic permutations to get
\begin{equation}
   \Lambda_{k k'}(\theta) =
   \Sp
   \left(
         \ckp \exp \left( \im \theta \, \Nop -\beta \, \Hop \right) \ckd
   \right).
\end{equation}
At this point, we insert the identity operator
$\exp ( \im \theta \, \Nop - \beta \, \Hop)
 \exp (-\im \theta \, \Nop + \beta \, \Hop)$
under the trace, in front of $\ckp$,
\begin{equation}
   \Lambda_{k k'}(\theta) =
   \Sp
   \left(
         \exp \left(  \im \theta \, \Nop - \beta \, \Hop \right)
         \exp \left( -\im \theta \, \Nop + \beta \, \Hop \right)
         \ckp \exp \left( \im \theta \, \Nop -\beta \, \Hop \right) \ckd
   \right),
\end{equation}
and apply the operator identity
\begin{equation}
   \exp \left( -\im \theta \, \Nop + \beta \, \Hop \right) \ckp
   \exp \left(  \im \theta \, \Nop - \beta \, \Hop \right)
   = \exp \left( \im \theta - \beta \epskp \right) \ckp
   \label{eq:identity}
\end{equation}
that proves valid for non-interacting particles. Indeed, given the
Hamiltonian $\Hop = \sum_k \epsk \ckd \ck$ both exponents in
$\exp \left( -\im \theta \, \Nop + \beta \, \Hop \right) \ckp
 \exp \left(  \im \theta \, \Nop - \beta \, \Hop \right)$ are found to
factorize while $\ckp$ commutes with each factor but the $k'$-th one.
Consequently, all factors appearing in the right exponent other than the
$k'$-th one can be shifted to the left so as to neutralize their inverse
counterparts. Hence, we are left with
\begin{equation}
   \exp \left( -\im \theta \, \Nop + \beta \, \Hop \right) \ckp
   \exp \left(  \im \theta \, \Nop - \beta \, \Hop \right)
   =
   \exp \left(  z \ckpd \ckp \right) \ckp \exp \left( -z \ckpd \ckp \right),
\end{equation}
where $z = \beta \epskp -\im \theta$. Differentiation of 
$u_{k'}(z) \equiv \exp \left(  z \ckpd \ckp \right) \ckp
\exp \left( -z \ckpd \ckp \right)$ with respect to $z$ yields a first-order
linear differential equation
\begin{align}
   \frac{\partial u_{k'}(z)}{\partial z} 
   & = \exp \left(  z \ckpd \ckp \right) \left[ \ckpd \ckp, \ckp \right]
       \exp \left( -z \ckpd \ckp \right) \nonumber \\
   & = -\exp \left( z \ckpd \ckp \right) \ckp \exp \left( -z \ckpd \ckp \right)
     = -u_{k'}(z)
\end{align}
to be solved with the boundary condition $u_{k'}(0) = c_{k'}$.
The trivial solution $u_{k'}(z) = \exp \left( -z \right) c_{k'}$ immediately
leads to the operator identity quoted in Eq.~(\ref{eq:identity}). \\
As a result, we obtain
\begin{equation}
   \Lambda_{k k'}(\theta) = \exp \left( \im \theta - \beta \epskp \right)
   \Sp
   \left(
         \exp \left(  \im \theta \, \Nop - \beta \, \Hop \right)
         \ckp \ckd.
   \right).
\end{equation}
Exploiting $\ckp \ckd = \delta^{\vphantom{\dagger}}_{k' k} + \xi \ckd \ckp$, we can
rewrite the above results as
\begin{align}
   \Lambda_{k k'}(\theta)
   & = \exp \left( \im \theta - \beta \epskp \right)
       \left[
             \delta_{k k'} G(\beta, \theta) + \xi \Sp
             \left(
                   \exp \left( \im \theta \Nop - \beta \, \Hop \right) \ckd \ckp
             \right)
       \right] \nonumber \\
   & = \exp \left( \im \theta - \beta \epskp \right)
       \left[
             \delta_{k k'} G(\beta, \theta) + \xi \Lambda_{k k'}(\theta)
       \right].
\end{align}
Hence, the trivial solution reads
\begin{equation}
   \Lambda_{k k'}(\theta) = \delta_{k' k}
   \frac{G(\beta, \theta)}
        {\exp \left( \beta \epsk \! - \im \theta \right) - \xi}
\end{equation}
In turn, the expression for the single-particle correlation function simplifies
to
\begin{equation}
   \Stwo(x, x') = \frac{1}{2 \pi Z_N} \sum_k \phi^*_k(x) \phi_{k}(x')
   \int_{-\pi}^{\pi} \!\! \d \theta \, \exp \left( -\im N \theta \right)
   \frac{G(\beta, \theta)}{\exp \left( \beta \epsk - \im \, \theta \right) - \xi}
\end{equation}
with the particle density $n(x) = \Stwo(x, x)$ emerging as a particular case.

\subsection{Pair correlation functions}
Introducing the pair correlation function (four point function) as
\begin{equation}
   \Sfour(x, x') = \fbraket{\psid(x) \, \psi(x) \, \psid(x') \, \psi(x')},
\end{equation}
we first write $\Sfour(x, x')$ as
\begin{align}
   \Sfour(x, x')
   & = \fbraket{\psid(x) \, \psi(x')} \delta(x - x') +
       \fbraket{\psid(x) \, \psid(x') \, \psi(x') \, \psi(x)} \nonumber \\
   & = n(x) \, \delta(x - x') +
       \fbraket{\psid(x) \, \psid(x') \, \psi(x') \, \psi(x)}.
\end{align}
Expanding again all field operators in the complete set $\{ \phik(x) \}$, we
obtain
\begin{equation}
   \fbraket{\psid(x) \, \psid(x') \, \psi(x') \, \psi(x)}
   = \sum_{k k'} \sum_{q q'} \phi^*_q(x) \, \phi^*_{q'}(x')
     \phi_{k'}(x') \, \phi_k(x) \fbraket{\cqd \, \cqpd \, \ckp \, \ck}.
\end{equation}
A lengthy but straightforward calculation involving another application of
the operator identity (\ref{eq:identity}) and the commutation relation
$[\ck, \cqd \, \cqpd] = \delta_{k q} \cqpd + \xi \delta_{k q'} \cqd$
leads to
\begin{multline}
   \fbraket{\cqd \, \cqpd \, \ckp \, \ck}
   = \frac{1}{2 \pi Z_N}
     \left(
           \xi \delta_{k q} \delta_{k' q'} + \delta_{k q'} \delta_{k' q}
     \right)
     \int_{-\pi}^{\pi} \!\! \d \theta \, \exp \left( -\im N \theta \right) \\
     \times
     \frac{G(\beta, \theta)}
          {\left(
                 \exp \left( \beta \epsk  - \im \theta \right) - \xi
           \right)
           \left(
                 \exp \left( \beta \epskp - \im \theta \right) - \xi
           \right)
          }.
   \label{eq:cqqkk}
\end{multline}
Correspondingly, the pair correlation function is given by
\begin{multline}
   \fbraket{\psid(x) \, \psid(x') \, \psi(x') \, \psi(x)}
   = \frac{\xi^N}{2 \pi Z_N} \sum_{k k'}
     \left(
           \xi |\phi_k(x)|^2 |\phi_{k'}(x')|^2 +
           \phi^*_{k'}(x) \, \phi^*_k(x') \phi_{k'}(x') \, \phi_k(x)
     \right)
     \\
     \times \int_{-\pi}^{\pi} \!\! \d \theta \,
     \exp \left( -\im N \theta \right)
     \frac{G(\beta, \theta)}
          {\left( \exp \left( \beta \epsk  - \im \theta \right) - \xi \right)
           \left( \exp \left( \beta \epskp - \im \theta \right) - \xi \right)
          }
\end{multline}

\appendix

\section{Principle of maximum entropy \label{sec:app:Entropy}}

Consider the density operator
\begin{equation}
   \rhoN = \sum_j p_{j,N} \fketbra{\psi_{j, N}}{\psi_{j, N}}
   \quad \text{with} \quad \sum_j p_{j, N} = 1,
\end{equation}
where $p_{j, N}$ is the probability that state $\fket{\psi_{j,N}}$ of the
$N-$particle subspace is occupied in thermal equilibrium, i.e. with a fixed
ensemble average $\UN$ for the energy~\cite{Jaynes1,Jaynes2,Grandy}:
\begin{equation}
   \UN = \Sp \left( \HNpart \rhoN \right) = \sum_j p_{j, N} \EjN.
\end{equation}
Maximizing the entropy
\begin{equation}
   \SN = -\kB \Sp \left( \rhoN \ln \rhoN \right)
       = -\kB \sum_j p_{j, N} \ln p_{j, N}
\end{equation}
imposes
\begin{equation}
  -\kB \frac{\partial }{\partial p_{j, N}} \sum_{j^{\prime }}
   p_{j^{\prime}, N}
   \left(
         \ln p_{j^{\prime}, N} + \alpha + \beta E_{\!j^{\prime}, N}
   \right)
   = 0,
\end{equation}
where $\alpha$ and $\beta $ are Lagrange multipliers for the normalization
and the energy condition, respectively. Hence $p_{j, N} = \frac{1}{\ZN}
\e^{-\beta \EjN}$ with $\ZN = \sum_j \e^{-\beta \EjN}$.
But, keeping in mind that $\ZN$ and $\beta$ are in fact functions of the
fixed value $\UN$, a more careful notation is introduced:
\begin{equation}
   p_{j, N}(\UN) = \frac{1}{\ZN(\UN)} \e^{-\beta(\UN) \EjN}, \qquad
   \ZN(\UN) = \sum_j \e^{-\beta(\UN) \EjN},
   \label{eq:Entropy:p&Z}
\end{equation}
and therefore
\begin{align}
   \UN
   & = \frac{1}{\ZN(\UN)} \sum_j \EjN \e^{-\beta(\UN) \EjN},
       \label{eq:Entropy:UNwith_beta(UN)} \\
   \SN(\UN
   & = \kB \ln \ZN(\UN) + \kB \beta(\UN) \UN.
       \label{eq:Entropy:SN}
\end{align}
where the equation for $\UN$ is a transcendental equation which determines
the Lagrange multiplier $\beta$, and hence the temperature $T$ if defined
as $\beta = 1 / (\kB T)$. The Helmholtz free energy
\begin{equation}
   \FN(\UN) \equiv \UN - \frac{1}{\kB \beta(\UN)} \SN(\UN)
   \label{eq:Entropy:FN_def}
\end{equation}
then becomes, as expected:
\begin{equation}
   \FN(\UN) \equiv -\frac{1}{\beta(\UN)} \ln \ZN(\UN).
   \label{eq:Entropy:FN_result}
\end{equation}
At first glance all these result are familiar. Less familiar is a
relationship between the entropy and the energy dependence of $\beta$.
Differentiating $\ZN$ with respect to $\UN$, one obtains
$\d \ln \ZN / \d \UN = -\UN \d \beta / \d U$, and hence
\begin{equation}
   \ln \ZN(\UN) - \ln \ZN(U_{0}) = -\UN \beta(\UN) + U_{0}
   \beta(U_0) + \int_{U_0}^{\UN} \!\! \d U \, \beta(U).
\end{equation}
Using~(\ref{eq:Entropy:UNwith_beta(UN)}--\ref{eq:Entropy:FN_def}), this
expression simplifies into
\begin{equation}
   \SN(\UN) - \SN(U_{0}) = \kB \int_{U_0}^{\UN} \!\! \d U \, \beta(U),
\end{equation}
showing how the entropy increases with increasing internal energy. In the
classical limit, with $\beta(U) = C / (\kB U)$ where $C$ is the specific
heat, the right hand side becomes $C\ln (\UN / U_0) = C \ln (T_N / T_{0})$,
consistent with the equipartition theorem.

So far, it was shown that the projection operator approach is consistent
with the standard interrelations between the thermodynamic quantities, all
derivable from the partition function $\ZN$ and the (given) internal energy
$\UN$. No attention was paid to the relevance of the projection operator
for the actual calculation of $\ZN$, which becomes now the main topic of
interest. Since $\mathsf{\hat{P}}_{N}^{2} = \PN$, and $\ZN$ can be
rewritten as $\ZN = \Sp (\e^{-\beta \HNpart})$, one obtains with little
effort from~(\ref{eq:Entropy:p&Z})
\begin{equation}
   \ZN(\UN) = \Sp \left( \e^{-\beta(\UN) \Hop} \PN \right),
   \label{eq:Entropy:ZN(UN)}
\end{equation}
regardless whether the particles are interacting or not. Without the
projection operator, this would be the grand canonical partition function,
for which the chemical potential is required as a Lagrange multiplier to
impose the \emph{average} number of particles. The present approach is bound
to work in the canonical ensemble with \emph{exactly} $N$ particles.

Until this point, a purist notation was followed, emphasizing that thermal
equilibrium means that the internal energy $\UN$ is fixed, and that a
Lagrange multiplier $\beta(\UN)$ is introduced
in~(\ref{eq:Entropy:UNwith_beta(UN)}) to fulfill this requirement
\cite{Jaynes1,Jaynes2,Grandy}. For practical purposes, this formal treatment is
less appropriate. It is much easier to consider $\beta$ as a function
argument
\begin{equation}
   \ZN(\beta) = \Sp \left( \e^{-\beta \Hop} \PN \right),
\end{equation}
which at the end of the calculations is connected to the internal
energy via
\begin{equation}
   \UN(\beta) = \frac{1}{\ZN(\beta)}
                \Sp \left( \Hop \e^{-\beta \Hop} \PN \right)
              = -\frac{\d}{\d \beta} \ln \left( \ZN(\beta) \right).
\end{equation}

\section{Two Euler identities\label{sec:app:Euler}}
Given two complex numbers $z$ and $q$, with $|q| < 1$, Leonhard Euler in
the 18th century derived (amongst a variety of other mathematical insights)
the following two identities:
\begin{align}
   \prod_{n = 0}^{\infty} \frac{1}{1 - z q^n}
   & = 1 + \sum_{n = 1}^{\infty} z^n \prod_{k = 1}^n \frac{1}{1 - q^k}
       \label{eq:euler1} \\
   \prod_{n = 0}^{\infty} \left( 1 - z q^n \right)
   & = 1 + \sum_{n = 1}^{\infty} z^n (-1)^n q^{n (n - 1) / 2}
       \prod_{k = 1}^n \frac{1}{1 - q^k},
       \label{eq:euler2}
\end{align}
In contemporary literature~\cite{Andrews,Bellman,GasperRahman,Berndt}, they
are usually obtained as a by-product of more general theorems on
$q$-products and $q$-series, which hinders a transparent derivation.
Therefore we propose an easily accessible proof, inspired by a strategy of
Berndt \cite{Berndt}.
Given a set of complex numbers $a, b, q, z$ with
$|z| < 1, |az| < 1, |bz| < 1, |q| < 1$, define a function
\begin{equation}
   f(z) = \prod_{n = 0}^{\infty} \frac{1 - a z q^n}{1 - b z q^n}
\end{equation}
and calculate $f(q z)$,
\begin{equation}
   f(qz)
   = \prod_{n = 0}^{\infty} \frac{1 - a z q^{n + 1}}{1 - b z q^{n + 1}}
   = \prod_{n = 1}^{\infty} \frac{1 - a z q^n}{1 - b z q^n}
   = \frac{1 - b z}{1 - a z}
     \prod_{n = 0}^{\infty} \frac{1 - a z q^n}{1 - b z q^n}
   = \frac{1 - b z}{1 - a z} f(z).
\end{equation}
The latter can conveniently be rewritten as
\begin{equation}
   f(z) - f(q z) = b z f(z) - a z f(q z).
   \label{eq:funceq}
\end{equation}
Since $f(z)$ is analytic wherever $|z| < 1$, we may assign a power series
to it:
\begin{equation}
   f(z) = \sum_{n = 0}^{\infty} C_n z^n,
   \label{eq:fseries}
\end{equation}
where $C_0 = f(0) = 1$ holds by construction of $f(z)$. \\
Substituting (\ref{eq:fseries}) into (\ref{eq:funceq}), we obtain
\begin{equation}
   \sum_{n = 0}^{\infty} C_n \left( 1 - q^n \right) z^n =
   \sum_{n = 0}^{\infty} C_n \left( b - a q^n \right) z^{n + 1}.
   \label{eq:fserieseq}
\end{equation}
Clearly, the $n = 0$ term in the left-hand side of Eq.~(\ref{eq:fserieseq})
vanishes, while its right-hand side may be rephrased by shifting the
summation index $n + 1 \to n$, yielding
\begin{equation}
   \sum_{n = 1}^{\infty} C_n \left( 1 - q^n \right) z^n =
   \sum_{n = 1}^{\infty} C_{n - 1} \left( b - a q^{n - 1} \right) z^n.
\end{equation}
Identification of the coefficients of $z^n$ then leads to the recurrence
relation
\begin{equation}
   C_n = \frac{b - a q^{n - 1}}{1 - q^n} \, C_{n - 1}, \qquad n \geqslant 1,
\end{equation}
which can be solved with the help of $C_0 = 1$ to get
\begin{equation}
   C_n = \prod_{k = 1}^n \frac{b - a q^{k - 1}}{1 - q^k},
\end{equation}
Filling $C_n$ out in (\ref{eq:fseries}) gives a generalization of the
well-known $q$-binomial theorem
\begin{equation}
   \prod_{n = 0}^{\infty} \frac{1 - a z q^n}{1 - b z q^n} =
   1 + \sum_{n = 1}^{\infty} z^n 
   \prod_{k = 1}^n \frac{b - a q^{k - 1}}{1 - q^k}
   \label{eq:qbinth}
\end{equation}
The Euler identities (\ref{eq:euler1}) and (\ref{eq:euler2}) 
emerge as special cases of (\ref{eq:qbinth}), corresponding respectively
to the cases $a = 0, b = 1$ and $a = 1, b = 0$.
 
\section*{References}
\bibliography{CanonicalEnsemble-MLB}

\end{document}